\documentclass{Interspeech2024}
\usepackage{amsmath}
\usepackage{amssymb}
\usepackage{mathtools}
\usepackage{amsthm}
\usepackage{multirow}
\usepackage{makecell}

\usepackage{hyperref}
\usepackage{cite}

\interspeechcameraready

\title{SimpleSpeech: Towards Simple and Efficient Text-to-Speech with Scalar Latent Transformer Diffusion Models}

\name{Dongchao}{Yang}
\name{Dingdong}{Wang}
\name{Haohan}{Guo}
\name{Xueyuan}{Chen}
\name{Xixin}{Wu}
\name{Helen}{Meng} 

\address{
  The Chinese University of Hong Kong}
\email{dcyang@se.cuhk.edu.hk} 


\begin{document}

\maketitle

\begin{abstract}
In this study, we propose a simple and efficient Non-Autoregressive (NAR) text-to-speech (TTS) system based on diffusion, named SimpleSpeech. Its simpleness shows in three aspects: (1) It can be trained on the speech-only dataset, without any alignment information; (2) It directly takes plain text as input and generates speech through an NAR way; (3) It tries to model speech in a finite and compact latent space, which alleviates the modeling difficulty of diffusion. More specifically, we propose a novel speech codec model (SQ-Codec) with scalar quantization, SQ-Codec effectively maps the complex speech signal into a finite and compact latent space, named scalar latent space. Benefits from SQ-Codec, we apply a novel transformer diffusion model in the scalar latent space of SQ-Codec. We train SimpleSpeech on 4k hours of a speech-only dataset, it shows natural prosody and voice cloning ability. Compared with previous large-scale TTS models, it presents significant speech quality and generation speed improvement. Demos are released.\footnote{https://simplespeech.github.io/simplespeechDemo/} \footnote{Accepted by Interspeech 2024. Code and pre-trained models will be released in the future.}
\end{abstract}

\section{Introduction}
Text-to-speech synthesis (TTS) aims to synthesize intelligible and natural speech given text, which has made great progress in the past years \cite{tecatron2,fastspeech2,cheng2023mrrl,yang2024instructtts,cheng2023ml,yang2022norespeech}. Most previous TTS systems trained on small-scale high-quality labeled speech datasets. In terms of model training, they rely on a relatively complicated pipeline (\textit{e.g.} prosody prediction, G2P conversion), and fine-grained alignment information (\textit{e.g.} phone-level duration) is needed.
However, the recently proposed approaches based on large-scale speech data significantly simplify the TTS system\cite{borsos2023audiolm,valle,speartts,uniaudio,make-a-voice}.
For instance, language model (LM) based TTS, \textit{e.g.}VALL-E \cite{valle} uses a pre-trained audio codec model \textit{e.g.} Encodec \cite{encodec} to map the speech signal into a sequence of discrete tokens, then an auto-regressive language model is trained to translate the phoneme into speech tokens. It can generate more expressive speech, but also is troubled by the slow and unstable inference. To address these issues, Non-autoregressive (NAR) models, \textit{e.g.} NaturalSpeech 2 \cite{ns2}, SoundStorm \cite{soundstorm}, and VoiceBox \cite{voicebox}, are proposed. They have higher inference speed and better stability, but with the cost of 1) relying on phoneme-acoustic alignment information, and 2) a more complicated training process. 

In this work, we propose a simple and efficient TTS system, SimpleSpeech, which does not rely on any alignment information and generates high-quality speech in a NAR way.\footnote{High-quality refers to naturalness, intelligibility, and fluency} To build such a TTS system, where we meet with the following research problems:
(1) how to use the large-scale speech-only dataset to train a TTS model without any alignment information \textit{e.g.} phoneme-level duration; (2) how to design a generative model that can generate high-quality speech in a NAR way. (3) how to solve the duration alignment problem without using a specific duration model when we train the NAR model;
The main contributions of this study are summarized as
follows: \\
(1) We demonstrate that large-scale unlabeled speech data can be used to build an NAR TTS system. \\
(2) We propose a novel generation model, a scalar latent transformer diffusion model, which models the speech data in a finite and compact latent space. More specifically, we propose a speech codec model (SQ-Codec) based on scalar quantization, which maps the complex speech signal into a finite and compact latent space, named scalar latent space. Then we apply the diffusion model in the scalar latent space of SQ-Codec.  \\
(3) We propose to use sentence duration instead of phone-level duration for the NAR-based TTS model. The sentence duration is used to determine the length of the target sequence, then an in-context conditioning strategy is introduced to learn the fine-grained alignment between the condition and target sequence implicitly. Compared to previous works that predict the duration of each phoneme, the sentence duration gives more diversity in the generation process. More importantly, the sentence duration is easy to obtain in both the training and inference stages. Please refer to Section \ref{sentence duration} finds more details. \\
(4) Extensive experimental results show the effectiveness of SimpleSpeech. We also conduct a lot of ablation studies to explore the effectiveness of each part in SimpleSpeech.
 
\begin{figure*}[h]
    \centering
    \includegraphics[width=\textwidth]{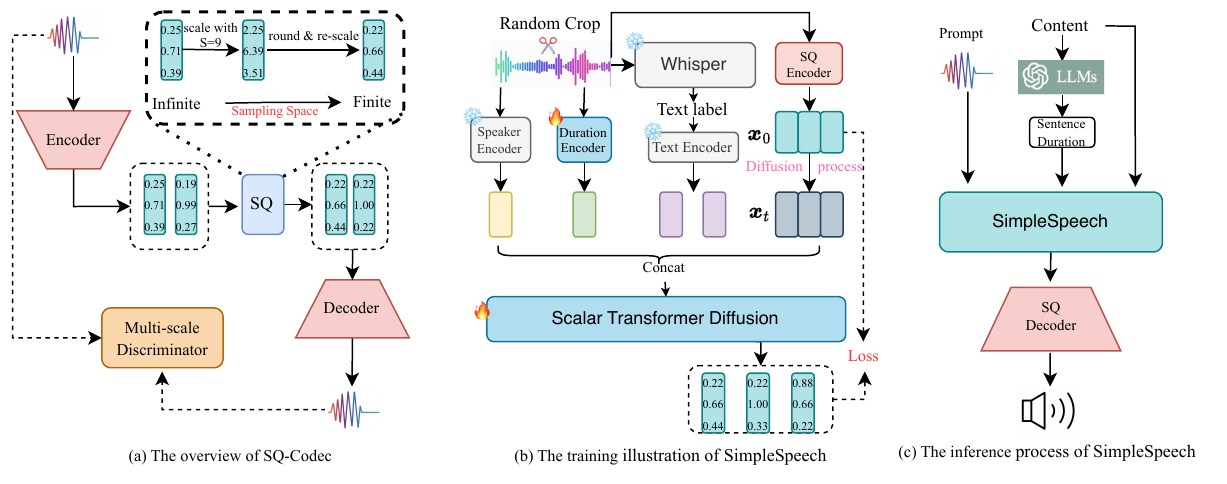}
    \vspace{-5mm}
    \caption{(a) shows the overview of the SQ-Codec. (b) and (c) give the illustration of SimpleSpeech in training and inference.} 
    \label{fig:uniaudio}
\end{figure*}

\section{SimpleSpeech} \label{sec:simplespeech}
The overall architecture of the SimpleSpeech framework is demonstrated in Figure 1 (b), which mainly consists of two parts: SQ-Codec and scalar transformer diffusion. The detailed design of each part will be introduced in this section. 

\subsection{Text Encoder and Speaker Encoder} \label{text encoder}
This work explores to use the large-scale speech-only datasets to train a TTS system. The first step is to obtain the text label for these speech samples. We propose to use the open available ASR model to get transcripts, Whisper-base model \cite{whisper} is used in this study. To take advantage of the large language model and simply the traditional TTS frontend, we use a pre-trained language model to extract the textual representations. Then we directly take these textual representations as the conditional information for TTS. To realize zero-shot voice cloning, we use the 1st layer of XLSR-53 \cite{conneau2020unsupervised} to extract global embedding to represent the speaker timbre. 

\subsection{Sentence duration} \label{sentence duration}
Previous works \cite{fastspeech2,ns2} try to model the phone-level duration in TTS. In general, a duration predictor is built to predict the duration of each phone. Training such modules increases the complexity of data pre-processing and training pipeline. In this study, we propose to model the sentence-level duration prior by using the in-context learning of LLMs (gpt-3.5-turbo is used). Our motivation is that LLMs can easily estimate how much time to read a sentence based on the number of words in the sentence and the prior knowledge. The prompt for ChatGPT can be found on the demo page. After we obtain the sentence-level duration, we let the model learn the alignment between words and latent features implicitly. Such a design will bring more diversity to speech synthesis. In the training stage, we can directly get the duration based on the length of the waveform. We follow Stable Audio \cite{stableaudio} uses a timing module to encode the duration into a global embedding. In the inference stage, the predicted duration by LLMs first determines the length of the noisy sequence, then input into the timing module.

\subsection{SQ-Codec} \label{sec:sq-codec}
Although residual vector quantization (RVQ) based audio codec models have shown effectiveness for audio compression, training a good codec model needs a lot of tricks and complicated loss design \cite{dac}. In this study, we propose to use scalar quantization \cite{balle2016end,mentzer2023finite} to replace the residual vector quantization in audio codec models, which can be trained with the reconstruction loss and adversarial loss without any training tricks. Furthermore, we also find that the scalar quantization effectively maps the complex speech signal into a finite and compact latent space, which is suitable for diffusion model (refer to Section \ref{sldm} and \ref{sec:ablation} for more details). Assuming $\boldsymbol{h} \in \mathcal{R}^{T*d}$ denotes the output features of Encoder in the codec model. $T$ and $d$ denote the number of frames and the dimension of each vector. For any vector $\boldsymbol{h}_i$, we use a parameter-free scalar quantization module to quantize $\boldsymbol{h}_i$ into a fixed scalar space:
\begin{equation}\label{codebook}
  \boldsymbol{h}_i = \text{torch.tanh}(\boldsymbol{h}_i), \quad
  \boldsymbol{s}_i = \text{torch.round}(\boldsymbol{h}_i*S)/S,
\end{equation}
where $S$ is a hyper-parameter that determines the scope of scalar space. To get gradients through the rounding operation, we use a straight-through estimator like VQ-VAE \cite{vqvae}. We can see that the scalar quantization first uses a \textit{tanh} function to map the value of features into $[-1, 1]$, then a \textit{round} operation further reduces the value of range into $\text{2*S+1}$ different numbers. We named such value domain as \textit{scalar latent space}. We note that previous works \cite{mentzer2023finite,yu2023language} also try to use scalar quantization as the image tokenizer for image generation. We claim that our implementation is different from theirs, and better adapts audio codec tasks in our experiments. \\
\textbf{Encoder and Decoder} Our encoder consists of 5 convolution blocks, each block includes 2 causal 1D-convolutional layers and one down-sample layer. The down-sample strides are set as $[2,2,4,4,5]$, resulting in 320 times down-sample along the time dimension.  The decoder mirrors the encoder and uses transposed convolutions instead of stride convolutions. \\
\textbf{Discriminator and Training Loss} Following \cite{uniaudio}, a multi-scale discriminator is used. The training loss of SQ-Codec consists of two parts: (1) reconstruction loss $\mathcal{L}_{rec}$, which includes time domain and frequency domain losses: the L1 loss between the reconstructed waveform and the original waveform and the MSE loss for the STFT spectrogram. (2) adversarial loss, which is calculated based on the results of the discriminator. 
\subsection{Scalar Latent Transformer Diffusion Models} \label{sldm}
Based on the previous discussion, we can map the speech data into a \textit{scalar latent space} based on our proposed SQ-Codec model. Inspired by the success of latent diffusion models \cite{ldm} in both image and audio generation \cite{make-an-audio,ns2}, we propose to model the speech data in the \textit{scalar latent space}. Our motivation is that the sampling space of \textit{scalar latent space} is simple because SQ effectively limits the value range of each element. The details of the scalar latent transformer diffusion model are as follows. \\
\textbf{Transformer-based diffusion backbone} 
U-Net backbone has been widely used in diffusion models, especially in the speech synthesis field \cite{ns2,diffvoice,diffwave,e3tts}. We also noted that many transformer-based diffusion models, such as DiT \cite{dit} and Sora \cite{sora} have been used in image/video generation. Inspired by the success of the transformer-based audio language models \cite{valle,speartts,uniaudio} and DiT, we propose a transformer-based diffusion backbone for speech synthesis. Specifically, a GPT2-like transformer backbone is used: 12 attention layers, 8 attention heads, and the model dimension is 768.  \\
\textbf{In-context conditioning} Inspired by LLMs and previous audio language models \cite{valle,speartts,uniaudio}, we simply append the features of time step $t$ and condition $c$ as the prefix sequence in the input sequence. Such a condition way allows us to use a standard GPT-like structure without modification. After the final block, we remove the conditioning sequence from the output sequence. \\
\textbf{Scalar latent diffusion} Latent diffusion models (LDM) have been demonstrated to fit complex data distributions, including VAE latent features \cite{ldm,diffvoice,make-an-audio,ns2}, Mel-spectrogram features \cite{diffgan-tts}, and waveform \cite{diffwave,e3tts}. We speculate that these data distributions are very complex because their search space is infinite. Instead, SQ-Codec provides a finite and compact \textit{scalar latent space}, thus we can consider modeling speech data in this space. Specifically, a network is trained to transfer the Gaussian distribution to the scalar latent space. We follow the training strategy of DDPM \cite{ddpm}, and the mean squared error (MSE) loss is used. To make sure the final output belongs to the scalar latent space, we use the scalar quantization (SQ) operation to limit the final prediction.
\begin{equation}\label{s ldm}
   \boldsymbol{\hat{x}}_0 = SQ(\theta(\boldsymbol{x}_T,T,\boldsymbol{c})) 
\end{equation}
where $\theta$ denotes the parameter of the neural network. $T$ denotes the timestep, $\boldsymbol{x}_T$ denotes the sampling features from the Gaussian distribution. $\boldsymbol{c}$ denotes the condition information. 
\section{Experiments} \label{sec:experiments-setting}
\subsection{Experimental setting}
\textbf{Dataset} We create a dataset with only 4k hours of unlabeled English speech data sampled from the Multilingual LibriSpeech (MLS) dataset \cite{mls}. For the evaluation set, we follow E3TTS \cite{e3tts}, evaluating the models on the LibriTTS test clean set \cite{zen2019libritts}. \\
\textbf{Model training} For the SQ-Codec model, we train it on the LibriTTS \cite{zen2019libritts} dataset. We set the sample rate as 16k Hz. Unless specifically stated, we set $S=9$ and $d=32$. We train the SQ-Codec with a learning rate of 2e-3, and Adam optimizer is used. For the diffusion, we train it on our selected MLS dataset. The AdamW optimizer with a learning rate of 1e-4 is used. \\
\textbf{Baselines} For codec models, we compare with publicly available audio codec models, including Encodec \cite{encodec}, DAC \cite{dac}, and HiFi-Codec \cite{yang2023hifi}. For TTS models, to establish a standard for comparison, our study employs five baseline models, including VALL-EX\cite{vallex}, Pheme TTS \cite{pheme}, X-TTS \footnote{https://github.com/Plachtaa/VALL-E-X, https://github.com/PolyAI-LDN/pheme, 
 https://github.com/coqui-ai/TTS}, E3TTS \cite{e3tts}, and NaturalSpeech 2 \cite{ns2}. The first three models are industry baselines with publicly available pre-trained models. E3TTS and NaturalSpeech 2 are not publicly available, we directly compare them with their released demos. E3TTS is the most related to us, which proposes to directly predict waveform using a diffusion model. However, E3TTS is trained on fixed length speech (\textit{e.g.} 10 seconds), such setting limits the flexibility of the model. Instead, SimpleSpeech supports the generation of a variable-length speech based on the sentence-level duration. \\
 \begin{table}[t]
    \centering
    \small
    \caption{Performance comparison between open-sourced audio codec models and ours. B denotes the Bitrate (kbps). H denotes the Hop-size. Evaluation is conducted on the LibriTTS test set.}
    \vspace{2mm}
    \scalebox{0.92}{
    \begin{tabular}{lccc|ccc}
    \toprule
    Model  &Size (M)  & H & B  & PESQ  & STOI & SSIM  \\
    \midrule
    Encodec & 14   & 75 & 12 & 3.76  & 0.90 & 0.72       \\
    DAC  & 70  & 50 & 8 & 3.99 & 0.95 & 0.85    \\
    HiFiCodec  & 60 & 50 &  2  & 3.24 &  0.88  & 0.72    \\
    \midrule
    Ours  & 5  & 50 & 8  & \textbf{4.16} & \textbf{0.95 }& \textbf{0.86} \\
    \bottomrule 
    \end{tabular}
    \label{tab:all-codec-comparison} }
\end{table}
\begin{table}[t]
    \centering
    \small
    \caption{The influence of parameters $S$ and $d$ for SQ-Codec.}
    \vspace{2mm}
    \begin{tabular}{lcc|ccc}
    \toprule
    Setting ID & $S$  &$d$    & PESQ  & STOI & SSIM  \\
    \midrule
      1 & 9 & 32    & 4.16 & 0.95 & 0.86       \\
      2 & 2  & 18    & 3.33 & 0.88 & 0.71    \\
      3 & 1  & 32    & 2.57 & 0.84 & 0.62    \\
    \bottomrule 
    \end{tabular}
    \vspace{-5mm}
    \label{tab:ablation-codec} 
\end{table}
\begin{table*}[t]
    \centering
    \small
    \caption{The comparison with previous zero-shot TTS models. MOS and SMOS are presented with 95\% confidence intervals.}
    \label{tab:zero-shot-tts}
    \scalebox{0.95}{
    \begin{tabular}{l|ccccc|cc}
    \toprule
    Model & MCD ($\downarrow$) & SIM ($\uparrow$) & WER ($\downarrow$) &DNSMOS ($\uparrow$) & Speed (seconds $\downarrow$) & MOS ($\uparrow$)   & SMOS ($\uparrow$)   \\
    \midrule
    GroundTruth & &-  & 2.7  & 3.82 &-  & 4.39 $\pm$ 0.07 & 4.26 $\pm$ 0.10 \\
    \midrule
    VALL-EX  & 8.16   & 0.946 & 5.6 &  3.75 & 34.5 & 4.07 $\pm$ 0.12 & \textbf{4.15 $\pm$ 0.20}   \\
    Pheme    & 8.56 & \textbf{0.958} & 4.6 & 3.70 & 3.3 & 3.89 $\pm$ 0.23  & 3.79 $\pm$ 0.29  \\
    X-TTS    & 9.93    & 0.949 & 5.6  & 3.80 & 2.77 & 4.19 $\pm$ 0.15 & 4.00 $\pm$ 0.13  \\
    SimpleSpeech (ours)  & \textbf{7.51}  & 0.956 & \textbf{3.3}  & \textbf{3.90}  & \textbf{1.6} & \textbf{4.45 $\pm$ 0.05} & {4.14} $\pm$ 0.12 \\ 
    \bottomrule
    \end{tabular} 
    \label{TTS}
    }
\end{table*}
\begin{table}[t]
    \centering
    \small
    \vspace{-2mm}
    \caption{The Small-Scale Subjective Test}
    \vspace{2mm}
    \begin{tabular}{lccc}
    \toprule
    Model   & E3TTS  & NaturalSpeech 2 & SimpleSpeech \\
    \midrule
      MOS & 4.32 $\pm$ 0.08 & 4.26 $\pm$ 0.08 &  4.38 $\pm$ 0.10   \\
      SMOS & 4.38 $\pm$ 0.17 & 4.38 $\pm$ 0.14 & 4.30 $\pm$ 0.11  \\
    \bottomrule 
    \end{tabular}
    \vspace{-3mm}
    \label{tab:small-scale} 
\end{table}
\noindent \textbf{Evaluation metrics}
For Codec models, we use the following metrics: Perceptual Evaluation of Speech Quality (PESQ), Short-Time Objective Intelligibility (STOI), and structural similarity index measure (SSIM). We use objective and subjective metrics to assess the zero-shot synthesis ability. For objective metrics, word error rate (WER) is used to measure the robustness of TTS models. Speaker Similarity (SIM) is used to access the clone ability. DNSMOS \cite{dnsmos} and Mel-Cepstral Distortion (MCD) are used to measure speech quality. To evaluate the generation speed, we calculate the average time of each model to generate 20 utterances in the same machine. For subjective metrics, we hire 30 listeners to conduct the MOS and Similar MOS (SMOS) evaluation. 
\subsection{Experimental Results} \label{sec:experiments}
\subsubsection{The performance of SQ-Codec model}
In this study, the codec model plays an important role in the TTS system, its reconstruction performance determines the upper limit of speech generation. Thus, we first conduct experiments to validate the reconstruction performance of our proposed SQ-Codec model. Table \ref{tab:all-codec-comparison} shows the performance comparison between previous audio codec models and our SQ-Codec. The results demonstrate that the SQ-Codec with the highest PESQ, STIO, and SSIM. Furthermore, the SQ-Codec model needs fewer model parameters. 

\subsubsection{The ablation study of SQ-Codec model}
In this part, we conduct an ablation study to explore the influence of parameters $S$ and $d$. When $S=1$, we follow previous work \cite{yu2023language} to quantize the value into -1 or 1. Although such a strategy gets good performance in image compression, we can see that it cannot work very well in audio compression. When we increase $S$ and $d$, we can see the performance improvement. 

\subsubsection{The performance of SimpleSpeech}
In this part, we conduct experiments to validate the performance of zero-shot speech synthesis. Table \ref{tab:zero-shot-tts} shows the results. \\
\textbf{Generation Quality} One of the problems in previous large-scale TTS is their poor speech quality. In this study, we use better audio codec models and a novel generation model, we can see that both the subjective and objective metrics show SimpleSpeech obtains better speech quality. From the MOS and DNSMOS scores, we can see the generated samples by SimpleSpeech even surpass the ground truth speech. \\
\textbf{Speaker Similarity} Speaker similarity is an important metric to access the performance of voice cloning in TTS. \footnote{The cosine similarity between the generated speech and prompt. The speaker encoder in Section \ref{text encoder} is used to extract global embedding.} Previous works have demonstrated that by increasing the speech data to a large scale (\textit{e.g.} 60k hours), the voice cloning ability will be improved. Although SimpleSpeech only trained on 4k hours of data, we can see that it also shows comparable performance to baselines trained with much more data. \\
\textbf{Robustness} We assess the robustness of TTS models by measuring the word error rate of generated speech (the  HuBERT model \cite{hsu2021hubert} is used). We can see that SimpleSpeech shows better robustness than previous works, \textit{e.g.} SimpleSpeech is more robust than VALL-EX. Similar to VALL-EX, SimpleSpeech does not need phone-level duration alignment information. \\
\textbf{Generation Speed} AR-based large TTS models, such as VALL-E and VALL-EX, try to generate the speech tokens in an AR way, which will cost a lot of time in the inference. In this work, our model first generates scalar latent features by NAR way, then these generated features can be used to recover the waveform using SQ-Codec's decoder. Furthermore, our SQ-Codec model is very small, so it further improves the generation speed. Our proposed model can get high-quality speech with only 100 DDPM steps, instead, E3TTS and NaturalSpeech 2 use 1000 steps and 150 steps to ensure sufficient generation quality. It also validates the efficiency of our proposed model.

\subsubsection{The ablation study of SimpleSpeech} \label{sec:ablation}
We conduct ablation studies to explore the influence of each part in SimpleSpeech. The experimental results are shown in Table \ref{tab:ablation}, and the qualitative findings are as follows. \\
\textbf{Text encoder} From Table \ref{tab:ablation}, we can see that using BERT \cite{bert} as the text encoder results in a decrease in robustness (WER from 3.3 to 19.2). We can see that ByT5 is critical for generating coherent speech. A potential explanation could be that ByT5 recognizes individual characters, whereas BERT relies on subword tokenization techniques (such as byte-pair encoding). Since words that are spelled similarly often sound alike, the ability to discern characters can enhance the model's ability.  \\
\textbf{Model Structure} Unet-based structure \cite{ldm,make-an-audio} have been widely used in diffusion models. In this study, we show that our proposed transformer-based diffusion models can bring better performance than the Unet-based model. \\
\textbf{Latent Space} We explore the influence of latent space, including VAE-latent space \cite{ldm}, and our proposed scalar latent space. For VAE-latent space, we follow Diffvoice \cite{diffvoice} to train a VAE model. The latent space dimension is set as 20 for both mean and variance. We can see that our model brings a significant improvement over previous VAE-based LDM methods. We also explore applying LDM to the latent space of the DAC codec (following \cite{ns2}), but we fail to train such a model. We conjecture that the latent dimension of DAC codec is 1024, it is hard for LDM to predict. We think that the dimension of latent space not only influences the reconstruction ability of the codec but also influences the performance of generation models.  \\
\textbf{Condition style} We explore the influence of different types of condition ways, including cross-attention-based and our proposed in-context (IC) condition. We can see that our proposed in-context condition brings better performance than the cross-attention condition when the condition features do not include explicit duration information. Similar to LM-based generation methods, in-context condition makes our model learn the duration information by itself. \\
\textbf{Data quantity} We compare the performance gap when training SimpleSpeech on LibriTTS (400h) and our selected MLS (4000h) datasets. We notice that scaling the data quantity can significantly improve the robustness of SimpleSpeech.
\begin{table}[t]   
    \centering
    \small
    \vspace{-2mm}
    \caption{The ablation study for each part in SimpleSpeech.}
    \vspace{2mm}
    \begin{tabular}{l|cccc}
    \toprule
    Types  & WER  & SIM  & MCD & DNSMOS  \\
    \midrule
    SimpleSpeech    & 3.3   & 0.956 & 7.51  & 3.90 \\
    \midrule
    \multicolumn{4}{l}{\textbf{Replacing the text encoder}} \\
    ByT5 $\rightarrow$ BERT       & 19.2   & 0.958 & 8.21  & 3.86        \\
    \midrule
    \multicolumn{4}{l}{\textbf{Replacing the diffusion backbone}} \\
    Transformer $\rightarrow$ Unet       & 5.75  & 0.949 & 8.16  & 3.86        \\
    \midrule
    \multicolumn{4}{l}{\textbf{Compared to VAE-based LDM}} \\
    SQ-Codec $\rightarrow$ VAE       & 5.1  & 0.952 & 9.04  & 3.49  \\
    \midrule
    \multicolumn{4}{l}{\textbf{Replacing the condition way}} \\
    IC $\rightarrow$ Cross-attention    & 9.6   & 0.958 & 8.29  & 3.90 \\
    \midrule
    \multicolumn{4}{l}{\textbf{Training Data}} \\
    MLS $\rightarrow$ LibriTTS       & 7.5   & 0.94 & 7.71  & 3.58       \\
    \bottomrule 
    \end{tabular} 
    \vspace{-3mm}
    \label{tab:ablation}
\end{table}
\section{Conclusion} \label{sec:conclusion}
In this study, we propose a simple and efficient TTS model named SimpleSpeech. SimpleSpeech can be trained on large-scale speech-only datasets without any additional data pre-processing, which significantly simplifies the efforts to train a TTS model. We propose a novel speech codec (SQ-Codec) based on scalar quantization. Then we apply a novel latent transformer diffusion model to the scalar latent space of SQ-Codec. Experimental results show the proposed model has better performance than the previous Unet-based latent diffusion model. Due to the NAR generation strategy, SimpleSpeech significantly improves the generation efficiency compared to previous LM-based TTS models. In the future, we will explore to scale the model and data size.

\section{Acknowledgement} \label{sec:acknowledgement}
This work gets help from my friends Songxiang Liu and Rongjie Huang. I would like to thank Songxiang, for his help to revise the manuscripts. I would like to thank Rongjie, for his discussion with me. This work cannot be done without their help. 

\bibliographystyle{IEEEtran}
\bibliography{mybib}

\end{document}